\documentclass[useAMS,usenatbib,fleqn]{mn2e}  
\usepackage{times}
\usepackage{graphicx}
\usepackage{amsmath}
\def\Msun{\hbox{$\rm\thinspace M_{\odot}$}}
\def\chandra{{\it Chandra}}
\def\xmm{{\it XMM-Newton}}

\def\xte{{\it RXTE}}
\def\xmm{{\it XMM-Newton}}

\def\gsim{\mathrel{\hbox{\rlap{\hbox{\lower4pt\hbox{$\sim$}}}\hbox{$>$}}}}
\def\lsim{\mathrel{\hbox{\rlap{\hbox{\lower4pt\hbox{$\sim$}}}\hbox{$<$}}}}
   \title[The rapid X-ray variability of NGC 4051]{The rapid X-ray variability of NGC 4051}

   \author[S. Vaughan et al.]{S. Vaughan$^{1}$\thanks{sav2@star.le.ac.uk},
        P. Uttley$^{2}$, K. A. Pounds$^{1}$, K. Nandra$^{3,4}$, T. E. Strohmayer$^{5}$\\
        $^{1}$X-ray \& Observational Astronomy Group, Department of Physics and Astronomy, University of Leicester, Leicester, LE1 7RH. \\
        $^{2}$School of Physics and Astronomy, University of
        Southampton, Southampton SO17 1BJ\\
	$^{3}$Max-Planck-Institut f\"{u}r extraterrestrische Physik (MPE), Giessenbachstr. 1, 85748 Garching, Germany\\
	$^{4}$Astrophysics Group, Imperial College London, Blackett Laboratory, Prince Consort Road, London SW7 2AZ\\
	$^{5}$Astrophysics Science Division, NASA's Goddard Space Flight Center, Greenbelt, MD 20771, USA
             }

\begin{document}

\date{Accepted 2011 January 10. Received 2011 January 7; in original form 2010 November 22}
 
\pagerange{\pageref{firstpage}--\pageref{lastpage}} \pubyear{2011}

\maketitle

\label{firstpage}

\begin{abstract}
  We present an analysis of the high frequency X-ray variability of NGC 4051 ($M_{\rm BH} \sim 1.7 \times 10^6$ \Msun) based on a series of \xmm\ 
  observations taken in 2009 with a total exposure of $\sim 570$ ks (EPIC pn).
  These data reveal the form of the power spectrum over frequencies from $10^{-4}$ Hz, below the
  previously detected power spectral break, to $\gsim 10^{-2}$ Hz, above the frequency of the innermost stable circular orbit (ISCO) 
  around the black hole ($\nu_{\rm ISCO} \sim 10^{-3} - 10^{-2}$ Hz, depending on the black hole spin parameter $j$). 
  This is equivalent to probing frequencies $\gsim 1$ kHz in a stellar mass ($M_{\rm BH} \sim 10$ \Msun) black hole binary system.
  The power spectrum is a featureless power law over the region of the expected  ISCO frequency, suggesting no strong enhancement 
  or change in the variability at the fastest orbital period in the system.  
  Despite the huge amplitude of the  flux variations between the observations (peak-to-peak factor of $\gsim 50$) the power spectrum appears to be 
  stationary in shape, and varies in amplitude at all observed frequencies following the previously established linear rms-flux relation.
  The rms-flux relation is offset in flux by a small and energy-dependent amount. The simplest interpretation of the offset is in terms of a very soft spectral 
  component that is practically constant (compared to the primary source of variability). One possible origin for this emission is a circumnuclear shock energised by
  a radiatively driven outflow from the central regions, and emitting via inverse-Compton scattering of the central engine's optical-UV continuum (as
  inferred from a separate analysis of the energy spectrum).
	A comparison with the power spectrum of a long \xmm\ observation taken in 2001 gives only weak evidence for non-stationarity in power spectral shape or amplitude.
	Despite being among the most precisely estimated power spectra for any active galaxy, we find no strong evidence for quasi-periodic oscillations (QPOs) 
	and determine an upper limit on the strength of a plausible QPO of $\lsim 2$ per cent rms in the $3 \times 10^{-3} - 0.1$ Hz range, and $\sim 5-10$ per cent 
	in the $10^{-4} - 3 \times 10^{-3}$ Hz range. 
  We compare these results to the known properties of accreting stellar mass black holes in X-ray binaries, with the aim further of developing a `black hole unification' scheme.
\end{abstract}


\maketitle
%

\section{Introduction}
\label{sect:intro}

  \begin{figure*}
   \centering
	\vspace{-6 cm}
   \includegraphics[width=13cm, angle=90]{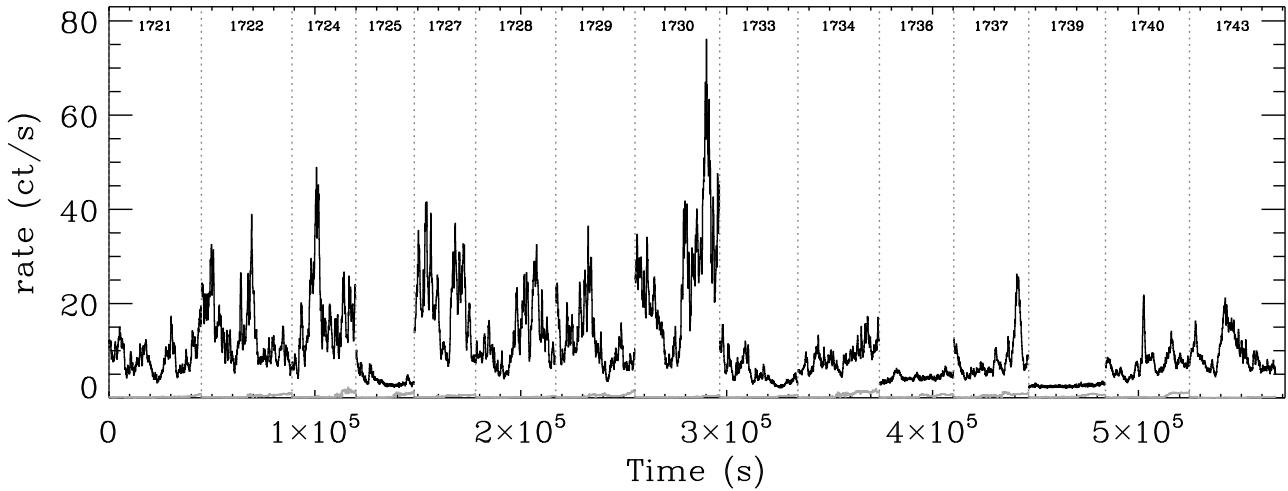}
      \caption{Time series of $0.2-10$ keV EPIC pn count rate (background-subtracted) from
      each of the $15$ observations taken in 2009. The bin size is
      $\Delta t = 100$ s. The $15$ separate $\sim 30-40$ ks observations, taken typically $~2-4$ days apart, have been concatenated in this plot, to allow easy comparison between them, with the end points marked by the dotted vertical line. The background count rate is also shown as a grey curve, but is almost always far weaker than the source. The labels above each section of data refer to the spacecraft revolutions during which the observations were made.}
         \label{fig:timeseries}
   \end{figure*}

If strong gravity dominates the dynamics of the inner accretion flows
around black holes then an elementary consequence is scale invariance:
many important aspects of accretion onto supermassive black holes ($M_{\rm BH} \gsim 10^6$~\Msun)
in Active Galactic Nuclei (AGN) should be fundamentally the same as
for stellar mass black holes ($M_{\rm BH} \sim 10$~\Msun) in Galactic
Black Hole systems (GBHs). (See e.g. \citealt{Shakura76} and \citealt{Mushotzky93}.)
Over the past few years several similarities
have been observed in the X-ray variability of nearby AGN and GBHs, supporting the 
idea of {\it black hole unification} \citep{Fender06}. In particular, the power spectrum (often power spectral density, PSD), and 
inter-band X-ray time lags appear to have similar frequency dependence and relative amplitudes in AGN and GBHs but with characteristic 
frequencies scaled (to first order) inversely with mass, $\nu \propto 1/M_{\rm BH}$ \citep[see e.g.][]{Lawrence87, Hayashida98, Edelson99, 
Uttley02, Vaughan03a, Markowitz03, Mchardy04, Done05, Mchardy06, Mchardy07, Gierlinski08, Arevalo08}.
Another property of the X-ray variability common to both types of system is a linear rms-flux relation, a scaling 
of the rms amplitude of variability with average source flux that appears to hold over a
very wide range of timescales \citep[][]{Uttley01, Vaughan03b, Gaskell04, Gleissner04,
Uttley05}. 

Arguably among the most important recent breakthroughs in the study of
stellar mass black holes was the discovery of quasi-periodic
oscillations (QPOs) at high frequencies ($\nu_{\rm Q} \gsim 100$~Hz)
in GBHs \citep[see][]{Remillard99, Strohmayer01, Remillard02, Remillard03, vanderklis06, Remillard06}.
These are the fastest
variations observed from GBHs and their high frequencies,
close to the Keplerian frequency of the
innermost stable circular orbit (ISCO; $\nu_{\rm ISCO}$), indicate an
origin in the central regions of the accretion flow.  High frequency QPOs (HF QPOs)
therefore carry information on the strongly curved spacetime close to the event horizon
and should, in principle, constrain the fundamental parameters of black holes: mass and spin.
As a product of strong gravity,
HF QPOs should also be present in AGN, but until recently there were no
robust detections, due mainly to the insufficient
length of previous \xmm\ observations \citep[see][]{Vaughan05b, Vaughan06}.
Recently a QPO was detected in the Seyfert
galaxy RE J1034+396 \citep{Gierlinski08a}.
This is a major discovery that adds further support to the general idea
that AGN are scaled-up GBHs, but this one detection provides limited diagnostic power
because it is so far unique and the other properties of the
source are not well understood \citep{Middleton09, Middleton10}

In parallel with these X-ray
advances, studies of transient radio jets from GBHs have led to the
development of a scheme that unifies jet production with accretion
state \citep{Fender04}. Relativistic jets
must be launched from  the region dominated by
strong-field gravity, and so scale invariance implies
this accretion state/jet unification scheme should extend to AGN
\citep{Heinz03}, a hypothesis supported by observations
\citep{Merloni03, Falcke04, Kording06}. From these results a new paradigm
is emerging in which accretion mode, X-ray spectrum, high-frequency
timing properties and jet production for both GBH ``states'' and AGN ``types''
may be unified into a single framework for the activity cycles of
accreting black holes.

Here we discuss the results obtained from a series of \xmm\ observations of the bright, 
highly-variable Seyfert 1 galaxy NGC 4051. These observations were designed to 
provide some of the best constraints to date on the high-frequency X-ray variability properties 
of any AGN and improve our understanding of black hole unification generally.
This particular target was chosen because of a combination of factors.
NGC 4051 \citep[redshift $z = 0.002336$, distance $D \approx 15$ Mpc;][]{Russell02}
has a relatively low and well-determined black hole mass, $M_{\rm BH} \approx 1.7 \pm 0.5 \times 10^6 ~ \Msun$ 
\citep{Denney09, Denney10}, displays strong, persistent X-ray flux and spectral variability
\citep[see e.g.][]{Lawrence87, Papadakis95, Mchardy04, Uttley04, Ponti06, Terashima09}, 
and a  soft X-ray spectrum rich in emission and absorption features 
\citep{Collinge01, Pounds04, Ogle04, Steenbrugge09}.


\section{Observations and data analysis}
\label{sect:obs}

NGC 4051 was observed by \xmm\ $15$ times over $45$ days during May-June 2009, and previously during 2001 and 2002 
(see table~\ref{tab:obs}).
The total duration of the useful data from the 2009 campaign is $\approx 572$ ks, giving 
$\sim 6 \times 10^6$ EPIC pn source counts (using {\tt PATTERN} $0-4$) or $\sim 9 \times 10^6$ EPIC 
source counts (using {\tt PATTERN} $0-12$ pn and MOS combined).

The raw data were processed from Observation Data Files (ODFs) following standard procedures 
using the \xmm\ Science Analysis System (SAS v10.0.2). The EPIC data were processed using the standard SAS processing chains to produce calibrated event lists. 
For each observation, source events were extracted from these using a $40$ arcsec circular region centred on the the target, 
and background events were extracted from a rectangular region on the same chip but not overlapping with the source region. (In order to obtain a precise background estimate the background region was $\sim 7$ times larger than the source region.)
Examination of the background time series showed the background to be relatively low and stable throughout 
most of the observations, with the exception of the final few ks of each observation where the background 
rate increased as the spacecraft approached the radiation belts at perigee (each of the observations 
occurred towards the end of a spacecraft revolution). The EPIC data were taken using the small window mode of each camera
and the medium blocking filter to reduce pile-up and optical loading effects, respectively. Nevertheless, during bright intervals NGC 4051 is sufficiently bright to
cause moderate pile-up, especially in the MOS, which can affect power spectrum estimation \citep{Tomsick04}. 
In order to mitigate any distorting effects all results are based on pn data PATTERNS $0-4$ only, but 
where possible checked for consistency with the MOS data. 
Response matrices were generated using {\tt RMFGEN v1.55.2} and ancillary responses were generated with {\tt ARFGEN v.1.77.4}. 

\begin{table}
\centering
\caption{Observation log. The columns list 
(1) the observation identifier, 
(2) the spacecraft revolution number, 
(3) the date of the start of the observation, 
(4) the observation duration.
}
\begin{tabular}{lrrr}
\hline\hline
Obs ID. &            & Obs. date  & Duration  \\
ID      & rev.       & (YYYY-MM-DD)  & (ks) \\
\hline
 0109141401  & 0263 & 2001-05-16  & 121958 \\
 0157560101  & 0541 & 2002-11-22  & 51866 \\
0606320101  & 1721 & 2009-05-03 & 45717 \\
0606320201  & 1722 & 2009-05-05 & 45645 \\
0606320301  & 1724 & 2009-05-09 & 45548 \\
0606320401  & 1725 & 2009-05-11 & 45447 \\
0606321301  & 1727 & 2009-05-15 & 32644 \\
0606321401  & 1728 & 2009-05-17 & 42433 \\
0606321501  & 1729 & 2009-05-19 & 41813 \\
0606321601  & 1730 & 2009-05-21 & 41936 \\
0606321701  & 1733 & 2009-05-27 & 44919 \\
0606321801  & 1734 & 2009-05-29 & 43726 \\
0606321901  & 1736 & 2009-06-02 & 44946 \\
0606322001  & 1737 & 2009-06-04 & 39756 \\
0606322101  & 1739 & 2009-06-08 & 43545 \\
0606322201  & 1740 & 2009-06-10 & 44453 \\
0606322301  & 1743 & 2009-06-16 & 42717 \\ 
\hline
\hline
\end{tabular}
\label{tab:obs}
\end{table}

Time series were extracted from the filtered events for the source and background regions using a range of time bin sizes and energy ranges. 
These were all corrected for exposure losses, as catalogued in the Good Time Information (GTI) header contained in each event file, except where less
than $30$ per cent of a bin was exposed, in which case the data were calculated by linear interpolation from the nearest `good' data either side 
and appropriate Poisson noise was added (this was needed for only $0.3$ per cent of the total exposure time from the pn, for $\Delta t = 5$ s time bins). The outputs of this process were regularly sampled and uninterrupted, exposure-corrected time series for source and background. The background time series was then subtracted from the corresponding source time series (after scaling 
by the ratio of the `good' extraction region areas). Figure \ref{fig:timeseries} shows the resulting time series from the $15$
separate 2009 observations.


\section{Power spectrum}
\label{sect:psd}

\begin{figure}
   \centering
   \includegraphics[width=6cm, angle=90]{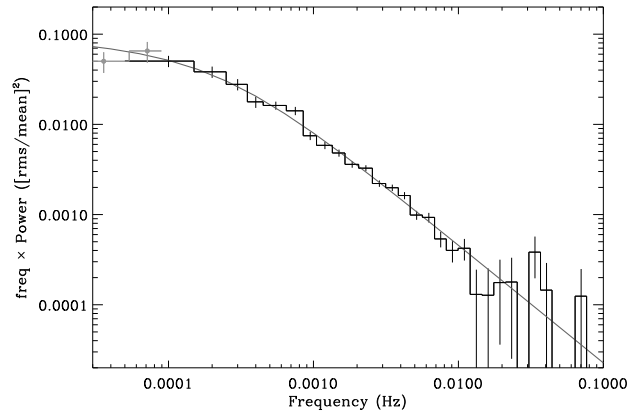}
      \caption{Power spectrum estimate (histogram) for the broad band ($0.2-10$ keV) EPIC pn data, computed using $10$ ks intervals at $\Delta t = 5$ bins, with
      best-fitting bending power law continuum model (solid curve). Also shown (grey circles) are two low frequency power density estimates computed using $28$ ks intervals.
      These lower frequency data were not used in the fitting (see text for details).}
         \label{fig:psd}
   \end{figure}

The power spectrum of the 2009 data was estimated using standard methods \citep{vanderklis89}. In 
particular, time series were extracted with binning $\Delta t = 5$ s and divided into uninterrupted
$10$ ks intervals. For each interval a periodogram was computed in units of relative power \citep{Miyamoto91, vanderklis95, Vaughan03b}, 
the  Poisson noise level\footnote{The expected power density due to measurement uncertainties in this case is 
$P_N = 2 \Delta T \langle \sigma^2 \rangle$, where $\Delta T$ is the sampling rate and $\langle \sigma^2 \rangle$ 
is the mean square error. See \citet{Vaughan03b}.} was subtracted, and then the periodgrams were averaged\footnote{
We have confirmed that the power spectral estimation and modelling was not significantly biased by sampling distortions -- commonly known as `aliasing' and `red noise leakage,' see \cite{vanderklis89, Uttley02, Vaughan03a}. Aliasing is minimal for contiguously binned data \citep{vanderklis89}, like those used here, and does not occur for the Poisson noise that dominates the highest frequencies probed. Leakage could in principle have been a problem, resulting from the finite duration of each observation, but in practice was not significant because the power spectrum was sufficiently flat ($\sim \nu^{-1}$) on timescales comparable to the observation length. This was confirmed using an analysis discussed in section \ref{sect:rms-flux}, footnote \ref{fn}.}.
Figure \ref{fig:psd} shows the resulting power spectrum estimate after rebinning in logarithmic frequency intervals.
The model fitting was performed using XSPEC v12.6.0 \citep{Arnaud96} using $\chi^2$ as the fit statistic, all data were binned to ensure that Gauss-normal statistics applied, and errors quoted correspond to $90$ per cent confidence limits (e.g. using a $\Delta \chi^2 = 2.71$ criterion for one parameter) unless otherwise stated.

The previous analysis of the power spectrum of NGC 4051 by \citet{Mchardy04} found a 
smoothly bending power law continuum provided an adequate description of the data
\begin{equation}
\label{eqn:bend}
  cont(\nu) = \frac{N \nu^{\alpha_{\rm low}}}{  1 + ( \nu / \nu_{\rm bend} )^{\alpha_{\rm low} - \alpha_{\rm high} } }   ,
\end{equation}
with indices of
$\alpha_{\rm low} \approx -1.1$ at low frequencies and $\alpha_{\rm high} \approx -2$ 
above $\nu_{\rm bend} \sim 8 \times 10^{-4}$ Hz. This model provided an excellent
fit to the 2009 power spectrum, as shown in Figure \ref{fig:psd}, with the lower frequency 
index fixed at $\alpha_{\rm low} = -1.1$ \citep[since it is much better constrained by the \xte\ data;][]{Mchardy04}. 
The best-fitting values for the free parameters of the model were $\alpha_{\rm high} = -2.31 \pm 0.08$ 
and $\nu_{\rm bend} = 2.3 \pm 1.0 \times 10^{-4}$ Hz. 
The fit statistic was $\chi^2 = 31.19$ with $27$ degrees of freedom (dof), giving a $p$-value of $0.26$. 
Allowing $\alpha_{\rm low}$ to vary provided no improvement in the fit ($\Delta \chi^2 < 2$) and the index itself was poorly 
constrained (with a $90$ per cent confidence interval of $[-0.25, -1.78]$), and
restricting $\alpha_{\rm low}$ to vary only within the range $[-1.2, -0.9]$ (the confidence interval determined by \cite{Mchardy04} from the \xte\ data) provided very little change in the confidence interval for $\nu_{\rm bend}$ compared to the fixed $\alpha_{\rm low} = -1.1$ model.
There is no obvious
indication, in either the overall fit statistic or the residual plot, of additional power
spectral components. We discuss this point in more detail in Section \ref{sect:qpo}.

A sharply broken power law model also provided a good fit, with $\chi^2 = 37.52$ for $27$ dof ($p=0.086$), slightly worse than the smoothly bending model.
The index below the break was again fixed to $\alpha_{\rm low} = -1.1$, and the best-fitting break frequency was $\nu_{\rm break} = 2.5 \pm 0.6 \times 10^{-4}$ Hz,
and a high frequency index of $\alpha_{\rm high} = -2.18 \pm 0.05$. These 
parameters, and the overall fit quality, are very similar to those of the bending power law model, indicating that the data are rather insensitive to 
the detailed shape of the change from flat to steep index. 
By contrast, the power spectrum is less well explained in terms of an exponentially cut-off power law, which gives 
$\chi^2 = 44.03$ for $26$ dof ($p = 0.015$), but only by allowing $\alpha_{\rm low} \approx -1.8$ which is inconsistent with the
long-term \xte\ results \citep{Mchardy04}. Assuming $\alpha_{\rm low} = 1.1$ gave an unacceptable fit ($\chi^2 = 217.81$ for $27$ dof; $p \ll 0.001$).

\begin{figure}
   \centering
   \includegraphics[width=6cm, angle=90]{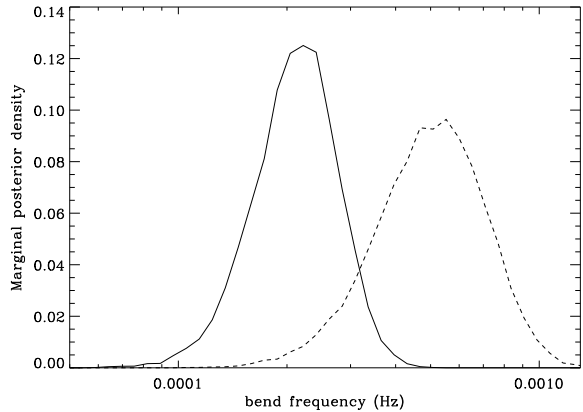}
      \caption{Marginal posterior probability density for the bend frequency in the bending power law model, 
      computed by MCMC (see text for details). That for the 2009 data is shown as a solid curve, and that for the 
      2001 data is shown as the dashed curve.}
         \label{fig:margin}
   \end{figure}

The model fitting was repeated using a Bayesian scheme by treating the likelihood function as $\sim \exp (-\chi^2/2)$ 
and assigning simple prior distributions to the parameters. See \cite{Vaughan10}, and references therein, for a brief 
introduction to these ideas. Uniform priors were assigned to $\alpha_{\rm high}$ and 
the overall normalisation, and to $\log \nu_{\rm bend}$ (equivalent to a $p(\nu) = 1/\nu$ Jeffreys' prior), although in practice
using uniform or Jeffreys' prior made no significant difference to the results. As expected, the parameter values at the 
posterior mode were almost identical to those giving the best-fit using the classical $\min \chi^2$ method.
But having the problem restated in Bayesian terms allowed for an exploration of the posterior distribution of the parameters, $p(\theta_C \mid \mathbf{x}^{\rm obs})$ with $\theta_C$ the parameters specifying the continuum and $\mathbf{x}^{\rm obs}$ the observed data, 
using a Markov Chain Monte Carlo (MCMC) scheme to randomly draw sets of parameters from 
the posterior\footnote{Specifically, five chains of length $5 \times 10^4$ were generated (after removing an equal length ``burn-in'' period), using
a Metropolis-Hastings algorithm with a multivariate Cauchy proposal distribution.  
Convergence of the chains was confirmed using the $\hat{R}$ statistic \citep{Gelman92, Gelman04}. The results from each chain were then combined to produce a 
large sample of posterior draws.}.
The marginal density for the bend frequency, $p(\nu_{\rm bend} \mid \mathbf{x}^{\rm obs})$, as computed using the MCMC data, is shown in Figure \ref{fig:margin}; 
the $90$ per cent credible interval on $\nu_{\rm bend}$ was $[1.3, 3.2] \times 10^{-4}$ Hz.

The limited durations of the individual 2009 observations makes it difficult to directly probe frequencies 
lower than $10^{-4}$ Hz. But for completeness the power spectrum was also estimated using longer uninterrupted intervals
($28$ ks duration, of which there were $15$ in 2009, one for each observation). However, the estimates of power density 
at the lowest frequencies were not normally distributed in this case, due to the small number
of estimates contributing to the average \citep{Papadakis93}. These data were not included in the power spectral fitting, 
but the lowest frequency points are shown in Figure \ref{fig:psd} and are clearly consistent with an extrapolation 
of the best-fitting model.

As a test for a possible high frequency cut-off in the power spectrum the bending power law model was modified by 
including an exponential cut-off (the {\tt highecut} model in {\tt XSPEC}). This provided only a very small improvement in the
overall fit statistic ($\Delta \chi^2 = 4.1$ for $2$ additional free parameters) with only poorly constrained parameters (cut-off 
frequency $\nu_{\rm cut} \le 8 \times 10^{-3}$ and $e$-folding frequency $\nu_{\rm fold} = 2 \pm 1 \times 10^{-2}$). 
The power spectrum estimated from the high flux data (see next section) has slightly higher signal-to-noise above $10$ mHz, 
but again shows no significant difference in the quality of the fit between the bending power law with and without an exponential
cut-off at high frequencies. 

\subsection{Energy dependence of the power spectrum}
\label{sect:epsd}

The dependence of the power spectrum on photon energy was investigated by dividing the data  into three broad energy bands -- $0.2$--$0.7$, $0.7$--$2$ and $2$--$10$ -- and
estimating and fitting the power spectrum for each of these. A simultaneous fit to all these spectra using the same continuum model (equation \ref{eqn:bend}) 
gave a rather poor fit ($\chi^2 = 118.9$ with $85$ dof, $p = 0.009$) with the parameters tied between the three spectra. But allowing  to 
be different for each energy band, while keeping $\nu_{\rm bend}$ tied between them, gave a much better fit ($\chi^2 = 98.5$ with $83$ dof, $p = 0.12$); 
the improvement is significant in an $F$-test, with $p < 0.001$, 
indicating the high frequency power spectrum is energy dependent. The slopes estimated for the low, medium and high energy bands were 
$2.31 \pm 0.07$, $2.24 \pm 0.09$ and $2.04 \pm 0.10$, respectively. The flattening of the power spectrum at higher energies is very similar to that 
detected by \cite{Mchardy04} from the 2001 data, and previously observed in several other Seyfert galaxies \citep[e.g.][]{Nandra01, Vaughan03}.
Allowing the bend frequency $\nu_{\rm bend}$ to be different for each energy band, but with $\alpha_{\rm high}$ tied between them, provided a
considerably worse fit ($\chi^2 = 108.0$ with $83$ dof) than the energy-dependent $\alpha_{\rm high}$ model.
Furthermore, allowing both $\nu_{\rm bend}$ and $\alpha_{\rm high}$ to differ between energy bands did not improve significantly on the variable $\alpha_{\rm high}$ fit 
($\chi^2 = 95.7$ with $81$ dof). In this case the best-fitting $\nu_{\rm bend}$ values for the low and medium energy bands were very similar while the hard band gave a 
somewhat lower value, but with a much larger confidence interval. We conclude there is strong evidence for a change in the high frequency slope with energy, but
the data do not resolve any difference in the location of the bend.


\section{rms-flux relation}
\label{sect:rms-flux}

  \begin{figure}
   \centering
   \includegraphics[width=6cm, angle=90]{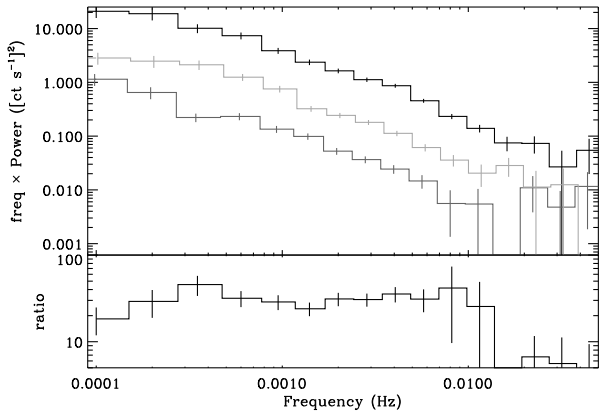}
      \caption{Power spectrum estimates computed as in Figure \ref{fig:psd} from $10$ ks intervals but averaged into three flux bins, with average ($0.2-10$ keV pn) count rates of $4.8$, $8.4$ and $17.4$ ct s$^{-1}$. The lower panel shows the ratio of highest to lowest flux power spectral estimates, which is consistent with a constant over two decades in frequency, indicating a strong increase in absolute power but no strong change in shape with average flux. The data were binned logarithmically in frequency, by a factor $1.4$, and the bin centres shifted slightly to prevent the error bars overlapping.}
         \label{fig:psd_flux}
   \end{figure}

The $15$ observations shown in Figure \ref{fig:timeseries} clearly show 
very different mean levels and amplitudes of variability, 
with noticeably stronger variations occurring preferentially during brighter 
intervals (e.g. rev1730), and the lowest flux intervals being extremely stable (e.g. rev1739).
This behaviour is exactly as expected for sources that obey the rms-flux relation
\citep{Uttley01, Uttley05}. 

In order to test this explicitly and in detail the power spectrum was
estimated in different flux bins. As before, the time series were divided into
uninterrupted $10$ ks intervals, the periodograms computed and the 
contribution of the Poisson noise subtracted, but using absolute units of power 
\citep[i.e. without normalising to the squared mean rate, see][]{Vaughan03b}. The intervals were sorted according to their mean flux, and the 
periodograms were then averaged in flux bins to provide estimates of the power
spectrum as a function of mean flux. Figure \ref{fig:psd_flux} shows the power 
spectral estimates when the data were divided into three flux bins. 
There is clearly a large difference in the overall 
amplitude (in absolute units) but little change in shape, as revealed by the ratio 
of bright and faint power spectra. 
Direct fitting of the high and low flux power spectra gives the same result: the best-fitting 
values for $\alpha_{\rm high}$ and $\nu_{\rm bend}$ estimated for the high flux data
fall within the $90$ per cent confidence intervals derived from  the low flux data.
The same effect was seen in Cygnus X-1 by \citet{Uttley01} (see their figure $2$), and
is consistent with there being a single rms-flux relation independent of the frequencies
used to compute the rms. 

\begin{figure}
   \centering
   \includegraphics[width=6.0cm, angle=90]{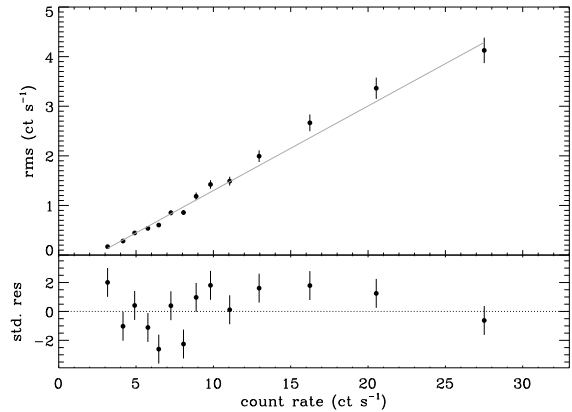}
      \caption{$0.2-10$ keV rms-flux relation. The rms is measured over the $1-10$
      mHz range (see text for details). The solid line marks the best-fitting linear model.	
      The lower panel shows the standardised residuals, i.e. (data -- model)/error.
      }
         \label{fig:rms-flux}
   \end{figure}

The detailed rms-flux relation was then examined explicitly by dividing the data into $1$ ks
uninterrupted intervals, computing a periodogram for each (again, in absolute units), subtracting the
Poisson noise from each, and then averaging the results in several flux bins\footnote{\label{fn}
We have checked that the power spectra estimated using short time series intervals (e.g. $1$ ks) are not significantly biased
by leakage \citep[see][and references therein]{Uttley02} in two different ways. First we compared the average power spectrum estimated from the real data using $1$ ks, $10$ ks and $28$ ks intervals and confirming they are all consistent. Secondly, by simulating data (using the best-fitting bending power law model) with the same sampling pattern as the real data, and estimating the bias between the power spectral estimate and the model.
} ($F_i$). The $1-10$ mHz rms, $\sigma_i$ was then 
estimated for each flux bin by integrating the corresponding power spectrum estimate over this frequency range. 
The resulting rms-flux data are shown in Figure \ref{fig:rms-flux}. (The uncertainties on the rms estimates were calculated based on the variance of the sum of the averaged periodograms, which are themselves calculated based
on the chi-square distribution of periodogram ordinates. The method is explained more fully in Heil, Vaughan \& Uttley in prep.)

\begin{figure}
   \centering
   \includegraphics[width=8.5cm, angle=0]{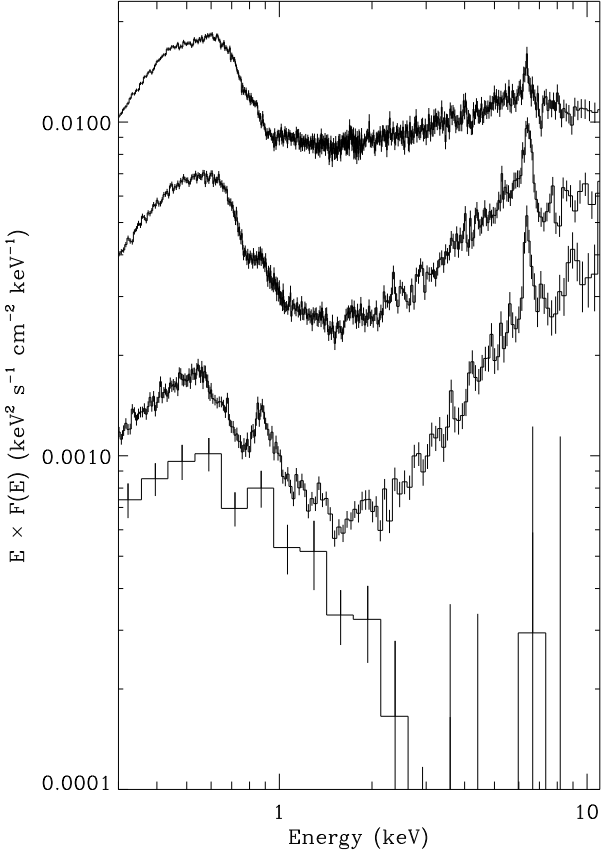}
      \caption{``Fluxed'' EPIC pn spectra shown in $E \times F(E)$ units. Moving from top to bottom 
      the spectra are: 
      (1) from a bright flux interval (rev1730), 
      (2) from a typical flux interval (rev1743)
      (3) from a faint interval (rev1739),
      (4) the offset of the rms-flux relation $C(E)$ (see text).
      Errors for the rms-flux 
      offset spectrum are the errors on the linear regression parameter. 
      }
         \label{fig:offset_spec}
   \end{figure}

The dependence of rms on flux is clearly very close to linear over the full flux range, as
previously noted in this source and accreting black holes
generally \citep[e.g.][Heil, Vaughan \& Uttley in prep.]{Uttley01, Uttley05}.
The best-fitting linear function -- $\sigma_i = k(F_i - C)$, where $k$ is the gradient and $C$ is the offset on the flux axis -- gave a rather high fit statistic ($\chi^2 = 30.05$ for $12$ dof) but the residuals show little  structure.
The best-fitting linear function has a negative offset in the rms axis, equivalent to a positive offset on the flux axis. 
The energy spectrum of this offset was computed by dividing the pn data into different energy bands ($14$ logarithmically spaced) and for each one calculating the rms-flux data and fitting with a linear model. The spectrum of the flux offset $C(E)$ is shown in Figure \ref{fig:offset_spec}, along with some representative pn spectra from bright (rev1730), intermediate (rev1743) and faint (rev1739) flux observations. The spectra were ``fluxed'' for display purposes, that is, they have been normalised to flux density units by dividing out the effective area curve (see Appendix \ref{sect:flux} for details). The spectrum of the offsets from the rms-flux relation is clearly soft: it is significantly positive in the lower energy bands, decreasing until consistent with zero intercept above $\sim 2$ keV. Most interestingly, it has a very similar shape and strength to the faint flux spectrum below $2$ keV.

\section{A search for quasi-periodic oscillations}
\label{sect:qpo}

The power spectral fitting discussed above provided no clear evidence for  narrow features, i.e. QPOs. 
Such features are usually described in terms of Lorentzian ``lines'' seen in the power spectrum in addition to 
the continuum. In order to assess the evidence for a QPO more rigorously we defined two competing hypotheses: 
the null hypothesis $H_0$ of a continuum power spectrum, and the alternative hypothesis $H_1$ which includes
an additional Lorentzian component.
\begin{equation}
  H_0 : P(\nu) = cont(\nu, \theta_C) 
\end{equation}
\begin{equation}
  H_1 : P(\nu) = cont(\nu, \theta_C) + qpo(\nu, \theta_Q).
\end{equation}
where the continuum model is defined above (equation \ref{eqn:bend}) and the QPO will be described by a Lorentzian profile
\begin{equation}
  qpo(\nu, \theta_Q) =  \frac{ 2 R^2 Q \nu / \pi} { \nu_{\rm Q}^2 + 4(\nu - \nu_{\rm Q})^2}.
\end{equation}
The parameters of the continuum are $\theta_C = \{ \alpha_{\rm high}, \nu_{\rm bend}, N \}$ and the 
QPO parameters are $\theta_Q = \{ \nu_{\rm Q}, Q, R \}$, where $\nu_{\rm Q}$ is the centroid frequency, 
$Q$ is the width parameter ($Q = \nu_{\rm Q} / \Delta \nu$, where $\Delta \nu$ is the Full Width at Half Maximum), and $R$ is the amplitude parameter 
(equal to the total rms in the limit of high $Q$). See \cite{Pottschmidt03} for details of this parameterisation.
In the present case we have assumed two different values, namely $Q=5$ and $Q = 10$, which are fairly typical values for high frequency QPOs in GBHs \citep[e.g.][]{Remillard02, Remillard03}. Using these two representative values rather than allowing $Q$ to be a free parameter made the QPO search and upper limit calculations (discussed later) much more efficient. 

One of the simplest ways to select between these two competing models is to assess the improvement in the $\chi^2$ fit statistic 
upon the addition of the QPO, i.e. the difference between the minimum $\chi^2$ for $H_0$ and $H_1$.
For a given dataset a large reduction (improvement) in the 
fit statistic between $H_0$ and $H_1$ is taken to be evidence in favour of a QPO. 
For the 2009 data, and the full $0.2-10$ keV energy band, the reduction in $\chi^2$ statistic was only $6.85$ (assuming $Q=10$, and a very similar value assuming $Q=5$), 
with best-fitting parameters $\nu_Q = 3.8 \pm 0.3$ mHz and $R = 1.2 \pm 0.5$ per cent. 
The modest improvement in the fit falls short of our threshold for a significant detection, as discussed below.
The QPO search was repeated for the three energy bands discussed above, with very similar results. The
$2$--$10$ keV band gave the largest improvement in the fit upon the addition of a QPO ($\Delta \chi^2 = 10.94$ assuming $Q=5$), 
but this still falls short of a detection in the $\alpha = 0.01$ test.

The lack of detection may itself be an interesting result if the observation was sufficiently sensitive,
and to assess this requires a calculation of the upper limit on the strength of a QPO.
The first step is to define a precise detection procedure and then calibrate its statistical `power' as a function
of QPO strength. That would allow us to 
determine the weakest QPOs that would be detected with reasonable probability.

\subsection{Detection procedure}

The search for QPOs in binned power spectral data is no different from the
search for ``lines'' in any other one dimensional spectrum, and suffers from 
the same statistical and computational challenges. See \cite{Freeman99}, \cite{Protassov02}, \cite{Park08} 
for elaboration of the issues.

The difference in the minimum chi-square statistic for each model was used as a test statistic. For a dataset $\mathbf{x}$ we compute
\begin{equation}
 T(\mathbf{x}) = \chi^2(\mathbf{x}, H_0) - \chi^2(\mathbf{x}, H_1).
\end{equation}
The nuisance parameters -- specifying the continuum and the QPO -- are eliminated by minimisation, i.e. $\chi^2(\mathbf{x}, H_0) = \min_{\theta_C} [\chi^2(\mathbf{x}, H_0, \theta_C)]$.
This test statistic is closely related to the Likelihood Ratio Test (LRT), as has been discussed by \cite{Protassov02}.
A large value of the test statistic (i.e. a large improvement in the fit upon adding a QPO to the model) is taken as evidence for a QPO.

The critical value of $T$ for detection, $T_{\rm crit}$, was calculated by a Monte Carlo method as discussed in Appendix \ref{sect:limit}.
In the present case we found $T_{\rm crit} = 12.28$ for $\alpha = 0.01$ (this is for $Q=10$, using $Q=5$ gave $13.05$ which is practically the same). In other words, assuming the continuum model to be true (with parameters represented by the posterior distribution $p(\theta_C \mid \mathbf{x}^{\rm obs})$ derived above from the real data), and no QPO, a reduction in the $\chi^2$ fit statistic of $T \ge 12.28$ will occur with probability $\alpha = 0.01$ (and have a posterior predictive $p$-value $\le 0.01$).

\subsection{Upper limit procedure}

In order to define an upper limit on the strength of possible QPOs 
we use the definition of an upper limit in terms of the power\footnote{Here we follow the notation used by \cite{Kashyap10} and use $\beta$ to denote the power of the test, i.e. the probability of meeting the detection criterion assuming the effect is real (e.g. there is a QPO present).} $\beta$ of a
detection procedure, as discussed very clearly in the recent paper by \citet{Kashyap10}. 
A QPO of a given frequency and strength $(\nu_Q, R)$ has a probability of detection $\beta(\nu_Q, R)$ using the above method (for fixed $\alpha$).

\begin{figure}
   \centering
   \includegraphics[width=6cm, angle=90]{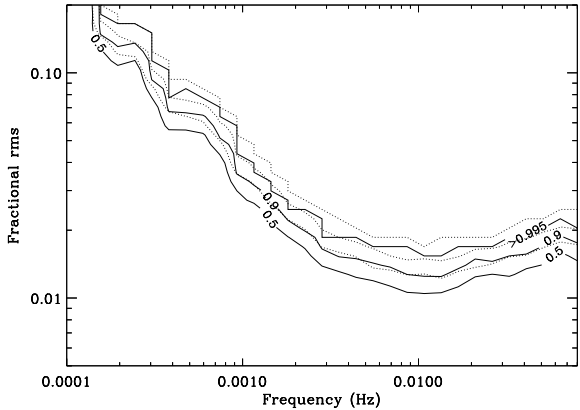}
      \caption{Upper limit on the strength of a QPO, assumed to have a Lorentzian spectrum, defined for an $\alpha=0.01$ test. 
		  The solid curve was calculated assuming a $Q=10$ QPO and the dotted curve was calculated assuming $Q=5$.
      In each case the three curves show (from bottom to top) the upper limits corresponding $\beta = 0.5, 0.9$ and $>0.995$ powers.
      The detection power at each point in the plane was computed using $200$ Monte Carlo simulations of data with a QPO present
      and recording the fraction of simulations that gave a detection (using a $\Delta \chi^2 \ge 12.28$ detection criterion for $Q=10$). See 
      section \ref{sect:qpo} and Appendix \ref{sect:limit} for details.}
         \label{fig:qpo_lim}
   \end{figure}

We used a Monte Carlo method, as discussed in Appendix \ref{sect:limit}, to estimate the detection probability $\beta(\nu_Q, R)$ for a range of 
different QPO locations and strengths ($\nu_Q$ and $R$ values). The estimate is simply the fraction of simulations (for each $\nu_Q$ and $R$) that gave a significant
detection defined by $T(\mathbf{x}^{\rm sim}) \ge T_{\rm crit}$. Contours of constant $\beta$ on the $\nu_Q$-$R$ plane define the upper limit on the
strength $R$ of a QPO as a function of frequency $\nu_Q$. Figure \ref{fig:qpo_lim} shows the contours for $\beta = 0.5$, $0.9$ and $>0.995$ assuming $Q=10$ (solid curves) and
$Q=5$ (dotted curves).
The middle of each set of curves represents the weakest QPO that could be present (at each frequency) and have a probability $\ge 0.9$ of being detected. This limit
extends below $R = 5$ per cent for $\nu \gsim 3 \times 10^{-4}$ Hz, and below $R = 2$ per cent for $\nu \gsim 3 \times 10^{-3}$ Hz. The shape and amplitude of this curve is similar
to that obtained using the simpler $\Delta \chi^2$ confidence contour method used previously by \cite{Vaughan05b}.

\section{Long term stationarity}
\label{sect:stationary}

\begin{figure}
   \centering
   \includegraphics[width=6cm, angle=90]{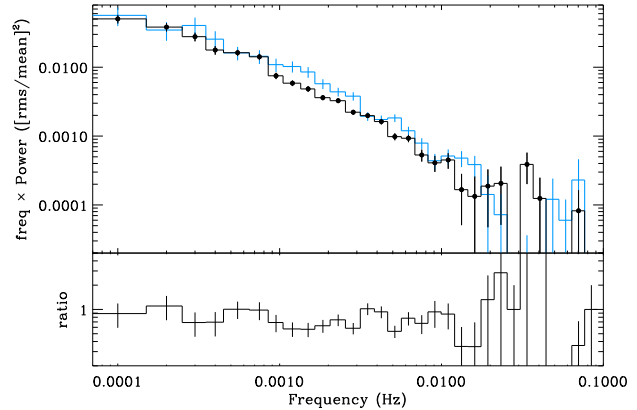}
      \caption{Comparison of 2009 and 2001 power spectrum estimates. Top panel: the estimated power spectra computed for the 2001 observation (blue histogram) and the combined 2009 observations (black; same as Figure \ref{fig:psd}). Bottom panel: ratio of the two estimates.}
         \label{fig:psd_new+old}
   \end{figure}

The power spectral estimation and fitting described in section \ref{sect:psd} was repeated for the 2001 data. 
Figure \ref{fig:psd_new+old} shows a comparison of the 2001 and 2009 power spectra which are in broad agreement
in shape and overall (fractional) amplitude.
The best-fitting bending power law model ($\chi^2 = 26.22$ for $27$ dof, $p=0.51$) had a slightly higher bend frequency ($\nu \approx 5.5 \times 10^{-4}$ Hz with a $90$ per cent confidence interval of $[3.0,9.0] \times 10^{-4}$ Hz, and a credible interval of $[2.7, 8.4]\times 10^{-4}$ Hz) and high-frequency index ($\alpha_{\rm high} = -2.48 \pm 0.16$) compared to the 2009 data. The marginal 
distribution of $\nu_{\rm bend}$ from both datasets are shown in Fig \ref{fig:margin}. These appear to suggest there was a slight change in the power spectral shape between the
two sets of observations. It is for this reason that we have not combined the 2001 and 2009 data to produce a composite power spectrum.

\section{Discussion and conclusions}
\label{sect:disco}

We have presented an analysis of the rapid X-ray flux variability of the low-mass Seyfert 1 galaxy NGC 4051 covering the range $10^{-4}$ to $\sim 2 \times 10^{-2}$ Hz, based on a series of $15$ \xmm\ observations taken during 2009. 
This is among the best  X-ray power spectra for any active galaxy, and the large range of fluxes sampled during the observations has allowed for an examination of the flux-dependence of the variability. Below we discuss the results and, where possible, relate them to the known behaviour of GBHs (for which we use a fiducial mass of $M_{\rm BH} = 10$ \Msun).

\subsection{Power spectral shape in relation to other sources}

The power spectrum of the variations is well described by a bending power law model that bends from an index of $\alpha_{\rm low} \approx -1.1$ to $\alpha_{\rm high} \approx -2.3$ around $\nu_{\rm bend} \sim 2 \times 10^{-4}$ Hz. 
This is fairly consistent with the \cite{Mchardy06} relation
\begin{equation}
  \log (T_{\rm b}) = 2.1 \log(M_6) - 0.98 \log (L_{44}) - 2.32,
\end{equation}  
where $T_{\rm b}$ is the break timescale\footnote{
Note that \cite{Mchardy06} defined $T_{\rm b}$ as the break frequency estimated by fitting power spectra with a sharply breaking power law model. This difference is of little consequence to the present discussion as we obtained very similar estimates for the characteristic timescale using smoothly bending and sharply broken models (section \ref{sect:psd}).
} in days ($\nu_{\rm bend} = 1.16 \times 10^{-5} / T_{\rm b}$ Hz), $M_{\rm BH} = M_6 10^6$ \Msun, and $L_{\rm Bol} = L_{44} 10^{44}$ erg s$^{-1}$ is the bolometric luminosity. The average $2-10$ keV X-ray 
flux during the 2009 observations was $F_{2-10} \approx 1.2 \times 10^{-11}$ erg s$^{-1}$ cm$^{-2}$ which gives $L_{2-10} = 3.5 \times 10^{41}$ 
erg s$^{-1}$  at an assumed distance of $15.7$ Mpc \citep{Russell02}, and $L_{\rm Bol} \sim  10^{43}$ \citep[assuming a $L_{\rm Bol}/L_{2-10} \sim 27$;][]{Elvis94}. The black hole mass is thought to be $M_{\rm BH} \sim 1.7 \pm 0.5 \times 10^6$ \Msun \citep{Denney09}, therefore $\dot{m}_{\rm Edd} \sim L_{\rm Bol} / L_{\rm Edd} \sim 0.05$.
Together these give a predicted bend frequency of $8 \times 10^{-5}$ Hz ($T_{\rm b} \sim 0.14$ day), within a factor $\sim 3$ of the observed value. In fact the estimated $\nu_{\rm bend}$ from the 2009 data is closer to the value expected from the \cite{Mchardy06} relation than that estimated by \cite{Mchardy04}, when calculated using the revised black hole mass estimate \citep{Denney09, Denney10}. 

The high frequency power spectrum above the bend rarely provides strong, unambiguous signatures of the accretion state of GBH systems, which tend to differ more at lower frequencies where features such as low frequency QPOs, flat-topped noise and broad Lorentzians occur \citep[see e.g.][]{vanderklis06, Mcclintock06, Belloni10}. 
\citet{Mchardy04} suggested NGC 4051 may be analogous to a soft state (particularly like that seen in Cygnus X-1) on the basis of the low frequency noise (a $nu^{-1}$ spectrum extending over two decades or more), the high frequency slope ($\alpha_{\rm high} < -2$) and the value of $\nu_{\rm bend}$ (which tends to be higher in the soft state, making the ratio $T_b/M_{\rm BH}$ smaller).  

For the 2009 data we found a very similar same high frequency shape and therefore is consistent with this identification.
Ignoring any $\dot{m}$ dependence and simply scaling frequencies inversely with mass ($\nu_{\rm bend} \propto 1/M_{\rm BH}$), the bend frequency for NGC 4051 correponds to 
$\sim 34$ Hz in a typical GBH. This is close to the $\nu_{\rm bend} \approx 23$ Hz estimated from a soft state observation of Cygnus X-1 by \citet{Mchardy04}.

The state analogy is supported by radio observations. At the highest available resolution the source is resolved 
into a core and two oppositely directed sources separated from the core by $\sim 0.4$ arcsec \citep{Ulvestad84, kukula95, Ho01, Giroletti09}.
This structure is similar to the core and hotspots commonly seen in Fanaroff-Riley II radio galaxies, although no radio jet has been detected in NGC 4051.
\cite{Jones11} recently used radio monitoring observations and found no evidence for strong variability of the core flux. 
Despite the presence of these radio sources, \cite{Jones11} showed that NGC 4051 falls below the `fundamental plane' connecting the radio and X-ray luminosities to black hole mass in radio loud AGN and hard-state GBHs \citep{Merloni03, Falcke04, Kording06}, i.e. it is less radio-loud than would be expected for a hard-state or radio-loud system \citep[see Fig. 14 of][]{Jones11}. 
This is consistent with the `soft state' interpretation of NGC 4051 mentioned above.

The \xmm\ data reveal the power spectrum of NGC 4051 up to frequencies as high as $\gsim 10^{-2}$ Hz (see Fig \ref{fig:psd} and \ref{fig:psd_flux}), a factor $\sim 40$ above the bend frequency $\nu_{\rm bend}$.
Assuming a black hole mass as above, the gravitational radius is $r_g = GM/c^2 \approx 2.5 \times 10^6$ km, and the period of the innermost stable circular orbit (ISCO)
should be $\sim 100-800$ s, depending on the dimensionless spin parameter, $j$ \citep[see e.g.][and references therein]{vanderklis06}. 
The power spectrum of NGC 4051 above $\nu_{\rm bend}$ shows no  indication of deviating from a power law up to the highest
frequencies (but see section \ref{sect:disco-qpo}), and in particular shows no signature of the ISCO period, which might be expected to produce an excess of variability power
due to Doppler boosting of ``hot spots'' \citep{Revnivtsev00, Sunyaev72}. 
There is no sign of the additional power spectral break seen at very high frequencies in the hard state of Cygnus X-1 by \cite{Revnivtsev00}\footnote{Above the break the power spectrum of NGC 4051 is sufficiently steep, $\alpha_{\rm high} < -2$, that the integrated power in the power spectra of $L_{\rm X}$ and $dL_{\rm X}/dt$ both converge at high frequencies. The rate of change of the luminosity may be
limited by the radiative efficiency of accretion \citep{Fabian79}. The power spectrum is therefore not required to become steeper at even higher frequencies.}.
Indeed, assuming linear scaling of timescales with black hole mass, we detect power at frequencies corresponding to $\sim 1.7$ kHz for a  GBH, far above the fastest variations yet detected from these sources \citep{Revnivtsev00, Strohmayer01}.

\subsection{Stationarity and the existence of ``states''}

The location of the power spectrum bend frequency is  lower than that reported by \cite{Mchardy04} from their analysis of the 2001 \xmm\ data. The $90$ per cent confidence intervals from these two analyses  -- $[5,12] \times 10^{-4}$ mHz and $[1.4,3.3] \times 10^{-4}$ Hz -- do not overlap, although some of this may be due to the interval reported by \cite{Mchardy04} being an underestimate \citep[see][]{Mueller09}. The difference between the two results may in part be due to differences in the 
data reduction and power spectrum estimation techniques; the re-analysis of the 2001 data presented here (section \ref{sect:stationary}) gives a lower $\nu_{\rm bend}$ estimate and shows the difference between the two observations to be only moderately statistically significant (compare also the Bayesian posterior distributions of Fig \ref{fig:margin}). 
If this is a real effect, it is different from the expected scaling of $\nu_{\rm bend}$ with accretion rate in the central region implied  by the \cite{Mchardy06} relation, since the X-ray luminosity in the month leading up to the 2009 observation (based on the \xte\ monitoring) was $\sim 20$ per cent higher than in the month leading up to the 2001 observations, but the bend frequency was $\sim 60$ per cent lower. 

The separation of $\sim 8$ years between the two observations corresponds to a separation of only $\sim 30$ minutes for a $10$ \Msun GBH; on these timescales GBH power spectra are approximately stationary except during very rapid state transitions. By the same linear scaling each of the $\sim 40$ ks \xmm\ observations is equivalent to $\sim 0.2$ s, and the $45$ day span of the observations covers the equivalent of only $\sim 20$ s for a typical GBH.
Unless NGC 4051 is drastically more non-stationary that most GBHs we might therefore expect all the \xmm\ observations to be representative of almost exactly the same state and variability properties.

During both the 2001 and 2009 observations the variability does, to a very good approximation, follow a single rms-flux relation over the full flux range (see Figure \ref{fig:rms-flux}) and the shape of the power spectrum remains roughly the same at different flux levels (see Figure \ref{fig:psd_flux}). The lack of variability during low flux intervals (e.g. rev1739) and the extreme variability during
high flux intervals (rev1730) are opposite extremes of this relation.
These apparently different behaviours are therefore not the result of distinct variability states\footnote{ 
Not in the sense normally applied to GBHs where different states are often characterised by distinct power spectral shapes and amplitudes as well as different energy spectra. See \citet{vanderklis06, Remillard06, Mcclintock06}.}, despite their very different energy spectra (see Figure \ref{fig:offset_spec}). (See also \citealt{Uttley03}.) The variations in energy spectral shape are also continuous, as discussed by Vaughan et al. (2010, in prep.).

\subsection{Absence of QPOs}
\label{sect:disco-qpo}

We found no strong  evidence (using a $\alpha = 0.01$ test) for QPOs (additional, narrow components) in the 2009 power spectrum, and we place limits of $R \lsim 2$ per cent rms on the strength of a QPO in the $\gsim 3 \times 10^{-3}$ Hz range, and $\lsim 5-10$ per cent over the $10^{-4} - 3 \times 10^{-3}$ Hz range. 
The $10^{-3} - 0.1$ Hz bandpass over which these data are sensitive to weak QPOs corresponds to $\gsim 170$ Hz for a GBH, 
where HF QPOs are often found, and the upper limit of $R \sim 2$ per cent on the QPO strength is comparable to the strength of typical HF QPOs 
\citep{Remillard03}. Such HF QPOs been detected in GBHs only in intermediate states \citep[particularly the soft intermediate state; ][]{Belloni10}, 
but are not always present (at least not at detectable levels), and when detected may be strongly energy-dependent.
The lack of detection in NGC 4051 therefore does not constitute strong evidence against an analogous state in this AGN. 

Despite not being formally significant in an $\alpha = 0.01$ test, the best-fitting narrow Lorentzian QPO component has a frequency of $\nu_Q = 4 \times 10^{-3}$ Hz and rms $R \sim 1$ per cent. This frequency is very close to that found previously by \cite{Vaughan05b} from fitting the power spectrum of the 2001 data. This may be a coincidence in the random sampling fluctuations around a smooth continuum power spectrum. But it is around this frequency that the signal-to-noise in the data is highest, and so model fitting will be most sensitive to small ($\sim 20$ per cent) deviations from the smooth power law model. The high frequency power law may be only an approximation to a more complex continuum power spectrum, the structure of which is on the edge of detectability in these data. 

It is well known that HF QPOs in GBHs can be strongly energy dependent, and frequently stronger at higher energies, and in fact we did find the higher energy band ($2$--$10$ keV) gave slightly stronger evidence for a possible QPO. But this could also indicate that this energy band contains a stronger contribution from a second, variable emission component with a different power spectrum from that which dominates the softer energy bands.

\subsection{A quasi-constant emission component}

The flux offsets in the energy-resolved rms-flux relations, $C(E)$, reveals the spectrum of a quasi-constant component (QC) of the emission.
A simple model comprising a single power law ($\Gamma = 3.06 \pm 0.13$) modified by Galactic absorption \citep{Elvis89}
gave a very good fit, with $\chi^2 = 10.65$ for $18$ dof. A Comptonising plasma model ({\tt compTT}) with optical
depth $\tau = 1$ and seed photon energy $\sim 50$ eV, which has an approximately power law shape, also gave a good fit ($\chi^2 = 9.45$ for $17$ dof).
The simplest interpretation is that this represents an emission component that does not vary at all during the
2009 observations; emission that remains if the highly variable emission component(s) drop to zero flux. 
We note that the \chandra\ data taken during an extended low flux period in 2001 and the \xmm\ observation taken during a low flux period in 2002 were previously
interpreted in terms of an unresolved  ($\lsim 100$ pc), non-varying soft spectral component and a strongly variable, harder spectral component \citep{Uttley03, Uttley04}. 
However, the QC emission is not required to be absolutely constant over the observations, it is only required to 
have a far lower (absolute) amplitude (at $\gsim 1$ mHz) and be weakly correlated (or uncorrelated) with the highly variable component 
that dominates brighter flux spectra (e.g. rev1730) and drives the rms-flux relation.

The spectral shape and absolute flux level of this component are similar to the lowest flux spectrum 
taken during 2009 (rev1739) below $\sim 2$ keV, as shown in Figure \ref{fig:offset_spec}. This strongly suggests that
most of the soft X-ray continuum emission observed during low flux periods is due to the soft QC component.
This may be the tail of the spectrum of inverse-Compton scattering of optical-UV continuum 
photons in shock-heated gas where the ionised outflow collides with the inter-stellar medium of the host galaxy, as recently 
predicted to occur in AGN by \citet{King10}. Indeed, the velocity-ionisation structure of the outflow as resolved by the RGS 
data provide further evidence to support this scenario \citep{Pounds10}.


\section*{Acknowledgements}
 
We thank the referee, Iossif Papadakis, for a thorough and constructive referee's report.
PU is supported by      
an STFC Advanced Fellowship, and funding from the European Community's       
Seventh Framework Programme (FP7/2007-2013) under grant agreement number     
ITN 215215 "Black Hole Universe".                    
This research has
made use of NASA's Astrophysics Data System Bibliographic Services, and the NASA/IPAC Extragalactic Data base
(NED) which is operated by the Jet Propulsion Laboratory, California
Institute of Technology, under contract with the National
Aeronautics and Space Administration.
This paper is based on
observations obtained with \xmm, an ESA science mission with instruments and
contributions directly funded by ESA Member States and the USA
(NASA).


\bibliographystyle{mn2e}
\bibliography{paper}

\appendix

\section{Construction of `fluxed' spectra}
\label{sect:flux}

X-ray spectra are conventionally given in terms of count rates per 
detector channel. These ``raw'' spectra are of limited use for
data visualisation because of the distorting effect of (i) 
the energy-dependent efficiency
of the telescope and detector (described by the effective area
function), and (ii) blurring by the line spread function
of the instrument (modelled by the redistribution matrix).
The true source spectrum $S(E)$ (in units of
photon s$^{-1}$ cm$^{-2}$ keV$^{-1}$) is related to the observed count rate in
channel $i$ by a Fredholm integral equation:
\begin{equation}
C(i) = \int R(i,E) A(E) S(E) dE 
\end{equation}
where $A(E)$ is the instrument effective area (cm$^{2}$) and
$R(i,E)$ gives the probability 
that a photon of energy $E$ will be recorded in channel $i$
(i.e. the redistribution function).
Here the observed spectrum $C(i)$ is assumed to be
background-subtracted, and the instrumental response defined by
$R(i,E)A(E)$ is assumed to be linear, that is, independent of
the value of $S(E)$ which  is a good approximation so long as pile-up
is negligible.
This integral equation can be approximated using a
discrete formulation in which the energy range is divided into $N$ bins $E_j$:
\begin{equation}
C_i = \sum_{j=1}^{N} R_{ij} A_j S_j
\label{eqn:fold}
\end{equation}
where $R_{ij}$ is the average and $A_j$ and $S_j$ are
the definite integrals of their continuous counterparts, 
over the energy range of each bin $E_j \rightarrow E_{j+1}$.

In general, methods that attempt to invert the above equations
to recover $S(E)$ from an observed spectrum $C_i$ are unreliable
and one must turn to the `forward fitting' method to build a model of
the spectrum \cite[see][]{Arnaud96}.  
However, it is often useful to view even an approximate
`fluxed' spectrum, largely free from the distorting effects of the instrument
efficiency, if only to give a better visual representation
of the underlying source spectrum. \cite{Nowak05} discussed a simple method 
to accomplish this in which the observed count spectrum
is  normalised by an effective area curve
that has been blurred using the redistribution function. 
\begin{equation}
\hat{S}_i = C_i / \sum_{j=1}^{N} R_{ij} A_j
\label{eqn:fluxed}
\end{equation}
This method 
matches the blurring of the effective area to the blurring of the
observed spectrum to give an estimate of the spectrum in true flux units (e.g. photon
s$^{-1}$ cm$^{-2}$) that is, to a large extent, free from the
distorting effects of the energy-dependent effective area.
This process is essentially equivalent to calibrating the
sensitivity using an observation of an ideal standard star
(with a perfectly flat spectrum)\footnote{
This
can be done trivially within {\tt XSPEC} by defining a constant model spectrum $M(E)=1.0$ 
(e.g. a power law with index $0$ and normalisation $1$) and plotting
the ``unfolded'' spectrum. 
For the special case of a constant model this is equivalent to
equation~\ref{eqn:fluxed}.
The `unfolded' plot shows the model in flux units ($=1$) multiplied by 
the ratio of the observed spectrum to the `folded'
model, i.e. $M(E_i) \times C_i / \sum_j R_{ij}A_j M_j = C_i / \sum_j
R_{ij}A_j$. the {\tt SAS} task {\tt efluxer} can also perform this transformation.}.
The resulting spectrum can easily be converted to
conventional flux density units (e.g. erg s$^{-1}$ cm$^{-2}$ 
keV$^{-1}$). 

It is crucial that the effective area function $A_j$
is blurred by the redistribution matrix.
The reason is that the effective
area function may contain abrupt changes in efficiency (e.g.
near  photoelectric edges in the detector-mirror system, such as the K-edges of O
and Si, or the M-edge of Au) that can be far sharper than the
resolving power of the detector, which means that normalising the observed (blurred)
spectrum by the exact (not blurred) effective area introduces very
strong spurious features into the spectrum near the edges. 
Blurring the effective area
curve matches the resolution to that of the data and suppresses this
effect. The result is a spectrum with the
energy-dependence of the detector/mirror efficiency removed (to a reasonable approximation), but 
which remains at the detector spectral resolution. We emphasise that this is not, even
approximately, a deconvolved spectrum.
It has the advantage over the popular `unfolding' method (in {\tt XSPEC}) of being independent of the
spectral model, and therefore providing a more objective visualisation `fluxed'
spectrum. 


\section{QPO detection and upper limits}
\label{sect:limit}

This appendix presents first an outline of the method used to search for QPOs in the power spectral data, 
and second how upper limits were derived based on the detection method.

\subsection{Calibrating the QPO detection procedure}

An $\alpha$-threshold detection procedure uses a test statistic $T$ to quantify evidence for a detection, and is calibrated 
such that a detection occurs when $T(\mathbf{x}) > T_{\rm crit}$ for some threshold value $T_{\rm crit}$ chosen to have a small probability $\alpha$ 
of occurring by chance when $H_0$ is true.
Specifically, the detection threshold is chosen such that $\Pr(T > T_{\rm crit} \mid H_0) \le \alpha$, where $\alpha$ is
the chance of a `false detection' (a `type I error' when $H_0$ is true). Commonly used values are $\alpha = 0.05$ and $0.01$.
But in order to calculate $T_{\rm crit}$ we need to know the  \emph{reference distribution} of the test statistic on the assumption  that the
null hypothesis is true, $p(T \mid H_0)$. In general this may be estimated by calculating the distribution of the test statistic from a large number of Monte Carlo simulations of data
generated assuming $H_0$ and distributed as expected for realistic data. 

The above would be valid if the null hypothesis was \emph{simple}, that is, did not contain any unknown (free) parameters.
In the present case the null hypothesis does involve parameters estimated using the observed data, $\theta_C$. 
We account for the 
uncertainties in their values by using random parameters drawn from the posterior distribution to generate the simulated data, using 
the MCMC discussed in section \ref{sect:psd}. This gives the distribution of the test statistic known as the 
posterior predictive distribution \citep[see][]{Rubin84, Meng94, Gelman96, Gelman04, Protassov02, Vaughan10}.
\begin{equation}
  p(T \mid H_0, \mathbf{x}^{\rm obs}) = \int p(T \mid H_0, \theta_C) p(\theta_C | \mathbf{x}^{\rm obs}) d\theta_C
\end{equation}
In order to determine $T_{\rm crit}$ this distribution was mapped out using Monte Carlo simulations as 
follows: 

\begin{itemize}

\item
repeat for $k=1,2,...,N$ simulated datasets:

\begin{enumerate}
    
    \item Draw a set of parameter values from the posterior $\theta_C^k$
    \item simulate data from $H_0$ with parameters $\theta_C^k$: $\mathbf{x}^k$
    \item fit $H_0$ to get $\chi^2(\mathbf{x}^k, H_0)$
    \item fit $H_1$ to get $\chi^2(\mathbf{x}^k, H_1)$
    \item compute $T^k$
\end{enumerate}

\item
 use the distribution of $T^k$ to define $T_{\rm crit}$
     for which $\Pr(T > T_{\rm crit} \mid  H_0) \le \alpha$

\end{itemize}

We used $N=3,000$ simulations to calibrate $T_{\rm crit}$ and found the critical value for 
an $\alpha = 0.01$ test was $T_{\rm crit} = 12.28$. (This was calculated using a $Q=10$ QPO, 
but using $Q=5$ gave a very similar result.)
for the data and models discussed in this paper.
In order to ensure the global minimum was found in each case (this can be tricky for $H_1$ which may produce multiple minima in $\chi^2$) we used the
robust automated fitting scheme discussed by \citet[][section 3.2.2]{Hurkett08}

\subsection{Upper limit procedure}

We wish to determine the upper limit $U$ on the QPO strength $R$, which
is the smallest value of $R$ such the probability of detecting the 
QPO (assuming it is present, i.e. $H_1$) is $\ge \beta_{ \rm min}$ for some
specific $\beta_{\rm min}$. 
See  \cite{Kashyap10} for a very clear introduction to the meaning and
calculation of upper limits in astronomy. 

In the present case we are searching for a QPO of unknown location (frequency) but the upper limit may vary as a function of this location --
because the ``background'' continuum level and the uncertainty on the spectrum vary as a function of frequency
-- so we must calculate the limit for a set of different QPO locations $\nu_Q$.
The probability of a detection, $\beta$, which is $1-$(probability of `type II error'), is:
\begin{equation}
  \beta(\nu_Q,R) = \Pr(T \ge T_{\rm crit} \mid H_1, \nu_Q, R)
\end{equation}
and the upper limit $U(\nu_Q)$ is the smallest value of $R$ which satisfies
\begin{equation}
  \beta(\nu_Q,R) \ge \beta_{\rm min} 
\end{equation}
for some chosen minimum power $\beta_{\rm min}$.

Given the $T_{\rm crit}$ defined above we can compute the upper limit
using Monte Carlo simulations of realistic data under
$H_1$. These can be used to map out the distribution $p(T \mid H_1, \nu_Q, R)$ 
for different $\nu_Q$ and $R$ values, and hence find $U(\nu_Q)$ that satisfies the 
above equations. In these terms the upper limit is the strength of the weakest QPO that would have a reasonable probability
$\beta \ge \beta_{\rm min}$, for some specific $\beta_{\rm min}$ such as $0.5$ or $0.9$. 

This method will, in general, be sensitive to the parameters of the continuum model, $\theta_C$, which of course we do not know exactly.
But, as before, we can average over the posterior distribution of these as derived from the data $p(\theta_C \mid \mathbf{x}^{\rm obs})$.
\begin{equation}
p(T \mid H_1, \nu_Q, R) = \int p(T \mid H_1, \nu_Q, R, \theta_C) p(\theta_C \mid \mathbf{x}^{\rm obs}) d\theta_C.
\end{equation}
Again, this can be done be tacking values from the posterior using the MCMC discussed in section \ref{sect:psd}.

The calculation of $\beta$ for a given pair of QPO parameters ($\nu_{Q}$ and $R$) was performed using a Monte Carlo scheme as follows.

\begin{itemize}
           
      \item  loop over $k=1,2,...,N$ simulated datasets
	    
	    \begin{enumerate}
   	    \item Draw a set of parameter values from the posterior $\theta_C^k$
	      \item  simulate data from $H_1$ with continuum parameters $\theta_C^k$ and QPO parameters ($\nu_{Q}, R$): $\mathbf{x}^k$
	      \item  fit $H_0$ to get $\chi^2(\mathbf{x}^k, H_0)$
	      \item  fit $H_1$ to get $\chi^2(\mathbf{x}^k, H_1)$
	      \item  compute $T_k$
	    \end{enumerate}

	    \item use distribution of $T^k$ to find
	    $\beta(\nu_{Q},R) = \Pr(T > T_{\rm crit} \mid H_1, \nu_{Q}, R)$ 

\end{itemize}

This process is repeated over a grid of QPO parameters, $\nu_{Q}$ and $R$, and at each frequency $\nu_{Q}$ 
the smallest $R$ for which $\beta(\nu_{Q},R) \ge \beta_{\rm min}$ is the upper limit $U(\nu_{Q})$.
For the present analysis we used $30$ logarithmically spaced values of QPO location $\nu_{Q}$, 
$40$ logarithmically spaced values of QPO strength $R$, and at each pair generated $N = 200$ simulations.
This is a sufficient number of simulations to accurately define contours of e.g. $\beta = 0.9$ on the $\nu_Q$-$R$ plane.

The detection procedure was calibrated using simulated data generated 
assuming $H_0$. But to evaluate $\beta$, and hence the upper limit, the data
were simulated assuming $H_1$. The former is concerned with the chance of 
spurious detection assuming no QPO present, the latter is concerned with the
chance of detection when a QPO is present.

\bsp

\label{lastpage}

\end{document}